\documentclass[aps,pre,twocolumn,groupedaddress]{revtex4-1}
\usepackage{graphicx}
\usepackage{dcolumn}
\usepackage{bm}
\usepackage{amsmath}
\usepackage{amssymb}
\usepackage{color}

\begin{document}

\title{Binding of curvature-inducing proteins onto tethered vesicles}

\author{Hiroshi Noguchi}
\email[]{noguchi@issp.u-tokyo.ac.jp}
\affiliation{Institute for Solid State Physics, University of Tokyo, Kashiwa, Chiba 277-8581, Japan}


\begin{abstract}
A tethered vesicle, which consists of a cylindrical membrane tube and a spherical vesicle,
is produced by a mechanical force that is experimentally imposed by optical tweezers and a micropipette.
This tethered vesicle is employed for examining the curvature sensing of curvature-inducing proteins.
In this study, we clarify how the binding of proteins with a laterally isotropic spontaneous curvature
senses and generates the membrane curvatures of the tethered vesicle using mean-field theory
and meshless membrane simulation.
The force-dependence curves of
the protein density in the membrane tube and the tube curvature are reflection symmetric and  point  symmetric, respectively,
from the force point, in which the tube has a sensing curvature.
The bending rigidity and spontaneous curvature of the bound proteins can be estimated from these force-dependence curves.
First-order transitions can occur between low and high protein densities in the tube
at both low and high force amplitudes. 
The simulation results of the homogeneous phases agree very well with the theoretical predictions. 
In addition,  beaded-necklace-like tubes with microphase separation are found in the simulation. 
\end{abstract}

\maketitle

\section{Introduction}

In living cells, numerous protein types work together to regulate biomembrane shapes~\cite{mcma05,suet14,joha15,bran13,hurl10,mcma11}.
Proteins are also involved in various dynamical processes, such as end/exocytosis and vesicle transport.
The Bin/Amphiphysin/Rvs (BAR) superfamily proteins bend the membrane along its axis
and generate cylindrical membrane tubes~\cite{mcma05,suet14,joha15,itoh06,mim12a}.
Other proteins, such as clathrin and coat protein complex (COPI and COPII), generate spherical buds~\cite{joha15,bran13,hurl10,mcma11}.
Thus, understanding the mechanism of these curvature generations is important.

These curvature-inducing proteins are known to sense the membrane curvature and are concentrated in membranes which have their preferred curvatures.
Various types of proteins can be examined using
a tethered vesicle pulled by optical tweezers and a micropipette~\cite{dimo14,baum11,sorr12,prev15,roux10,rosh17,alla20}.
With increasing force or length, the vesicle first deforms into a lemon shape and subsequently 
forms a narrow membrane tube (tether) protruding from a spherical vesicle~\cite{hota99,inab05,wu19}.
Moreover, an elongational force can be produced by the growth of protein filaments {\it in vitro}~\cite{hota99,gavr21} and {\it in vivo}~\cite{svit18,gall20}.
Curvature-inducing proteins typically bind more onto the membrane tube than the remaining spherical component.
BAR proteins~\cite{baum11,sorr12,prev15}, dynamin~\cite{roux10}, and G-protein coupled receptors~\cite{rosh17} have been reported to exhibit curvature sensing.

The aim of this study is to understand the curvature sensing and generation of
the curvature-inducing proteins with an isotropically spontaneous curvature on a tethered vesicle.
We employ mean-field theory and meshless membrane simulation.
In mean-field theory, a simplified geometry is considered for vesicles.
Previously, we used a vesicle consisting of many spherical components to study budding~\cite{nogu21a}.
Here, we apply the same scheme to the tethered vesicle.
Although mean-field theories have been used to analyze the experimental results of tethered vesicles~\cite{prev15,rosh17},
they have been applied in narrow ranges of the parameters,
and the curvature of the spherical component has not been considered.
Here, we systematically investigate the protein binding onto the tethered vesicle over a wide range of parameters.

Several types of membrane models have been developed for coarse-grained simulations~\cite{muel06,vent06,nogu09}.
For a large-scale simulation, we developed two types of meshless membrane models~\cite{nogu06,shib11},
in which membrane particles self-assemble into a single-layer membrane
and the mechanical properties can be varied over a wide range.
Here, we employ a spin meshless membrane model~\cite{shib11}, since 
it can vary the spontaneous curvature and
has been applied to membrane deformation by curvature-inducing proteins with an isotropic spontaneous curvature~\cite{nogu16a,gout21} and
 with an anisotropic spontaneous curvature~\cite{nogu14,nogu15b,nogu16,nogu17,nogu17a,nogu19a}, as well as topological changes of membranes~\cite{nogu19,nogu19c}.
In mean-field theory, we assume a uniform distribution of the bound proteins in each membrane component.
However, phase separation has been obtained in simulations of membrane tubes~\cite{nogu16a} and flat membranes~\cite{gout21}.
We clarify where the phase separation of the bound proteins occurs in the membrane tubes under a constant external force
and where the assumption of uniform protein density is valid.

The mean-field theory of the tethered vesicle is described in Sec.~\ref{sec:theory}.
Simulations of membrane tubes are described in Sec.~\ref{sec:sim}.
The simulation results are compared with the theoretical results in Sec.~\ref{sec:sres}.
Finally, a summary and discussion are presented in Sec.~\ref{sec:sum}.

\section{Tethered vesicle} \label{sec:theory}
\subsection{Mean-field theory}

\begin{figure}
\includegraphics[width=7cm]{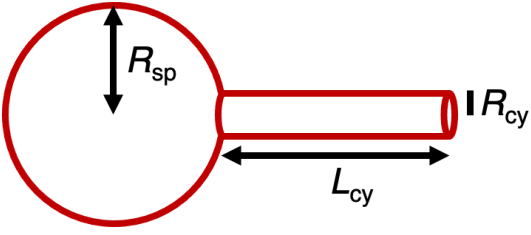}
\caption{
Simplified geometry of a tethered vesicle.
The vesicle consists of a sphere with a radius $R_{\rm sp}$ 
and a cylinder with a radius $R_{\rm cy}$ and length  $R_{\rm cy}$.
}
\label{fig:cat}
\end{figure}

A vesicle consists of a sphere with radius $R_{\rm sp}$ and
a cylinder with radius $R_{\rm cy}$ and length $L_{\rm cy}$ as depicted in Fig.~\ref{fig:cat}.
Here, a thin tube of $R_{\rm cy} \ll R_{\rm sp}$ is considered, 
such that the end and foot parts of the cylindrical tube are neglected.
This geometry was previously used in Ref.~\citenum{smit04}.
In experiments,
the end shape of the tube can be strongly affected by force-imposing methods such as optical tweezers,
whereas the foot has a catenoid-like shape with a low mean-curvature~\cite{powe02}.

The total surface area $A$ and volume $V$ are given as follows:
\begin{eqnarray}
  A &=& 4\pi R_0^2 = 4\pi R_{\rm sp}^2 + 2\pi R_{\rm cy}L_{\rm cy}, \label{eq:A} \\
  V  &=& \frac{4\pi}{3} R_{\rm sp}^3 + \pi R_{\rm cy}^2L_{\rm cy}, \label{eq:V} 
\end{eqnarray} 
where $R_0$ is the radius of a sphere with the same surface area.
From eqn~(\ref{eq:A}) and (\ref{eq:V}),
the radius and length of the membrane tube can be expressed as follows:
\begin{eqnarray}  \label{eq:a1}
  \frac{R_{\rm cy}L_{\rm cy}}{R_0^2} &=&  2(1- r^2), \\
\frac{R_{\rm cy}}{R_0} &=&  \frac{2(v_{\rm r}-r^3)}{3(1-r^2)},
\end{eqnarray} 
where $r = R_{\rm sp}/R_0$ and reduced volume $v_{\rm r} = V/(4\pi R_0^3/3)$.

Curvature-inducing proteins bind to the membrane, depending on the local membrane curvature.
The bending free energy $F_{\rm cv}$ is given by
\begin{equation}\label{eq:F0}
F_{\rm cv} = \int dA \ \Big\{ 2\kappa_{\rm d}H^2(1-\phi)
  +  \frac{\kappa_{\rm p}}{2}(2H-C_0)^2\phi   \Big\},
\end{equation}
where $H$ is the mean curvature of the membrane ($H=(C_1+C_2)/2$, where $C_1$ and $C_2$ are the principal curvatures).
The first and second terms represent the bending energy of the bare (unbound) and protein-bound membranes, respectively,
and $\phi$ is the protein density ($\phi=1$ at the maximum coverage). 
The unbound membrane has bending rigidity $\kappa_{\rm d}$ and zero spontaneous curvature;
the bound membrane has a larger bending rigidity $\kappa_{\rm p}$ and spontaneous curvature $C_0$.
In this study, the effects of the saddle-splay modulus $\bar{\kappa}$ are not considered.
Although  $\bar{\kappa}$ may depend on the protein density, 
the cylindrical membrane has zero Gaussian curvature ($C_1C_2$) such that
it has no influence in the middle of the tube.
Small effects may appear through the neglected regions (the end and foot of the membrane tube).
More general aspects of the protein-binding model are discussed in Ref.~\citenum{nogu21a}.

To induce a tether (thin membrane tube), an external force $f_{\rm ex}$ is imposed.
Here, it is assumed that the proteins are  homogeneously distributed in each membrane component 
(the spherical or cylindrical component).
The subscripts sp and cy represent the quantities of spherical and cylindrical components of the vesicle, respectively.
The free energy $F$ of the tethered vesicle is given by
\begin{eqnarray}
F &=&  F_{\rm sp} + F_{\rm cy} - f_{\rm ex}(L_{\rm cy}+2R_{\rm sp}), \\
\frac{F_{\rm sp}}{4\pi} &=&  \label{eq:FSP0}
 2(\kappa_{\rm dif}\phi_{\rm sp}+\kappa_{\rm d}) -2\kappa_{\rm p}C_0R_0r\phi_{\rm sp} \\ \nonumber
&& + R_0^2r^2(\sigma_{\rm p}\phi_{\rm sp} + b \phi_{\rm sp}^2) \\ \nonumber
&&+  \frac{ k_{\rm B}TR_0^2 r^2}{a_{\rm p}}[\phi_{\rm sp} \ln(\phi_{\rm sp}) +  (1-\phi_{\rm sp}) \ln(1-\phi_{\rm sp}) ],  \\
\frac{F_{\rm cy}}{\pi} &=& 
(\kappa_{\rm dif}\phi_{\rm cy}+\kappa_{\rm d})\frac{L_{\rm cy}}{R_{\rm cy}} - 2\kappa_{\rm p}C_0L_{\rm cy}\phi_{\rm cy} \\ \nonumber
&& + 2R_{\rm cy}L_{\rm cy}\Big\{(\sigma_{\rm p}\phi_{\rm cy} + b \phi_{\rm cy}^2) \\ \nonumber
&&+  \frac{k_{\rm B}T}{a_{\rm p}}[\phi_{\rm cy} \ln(\phi_{\rm cy}) +  (1-\phi_{\rm cy}) \ln(1-\phi_{\rm cy}) ]\Big\} \\
&=&  \label{eq:FCY0}
(\kappa_{\rm dif}\phi_{\rm cy}+\kappa_{\rm d})\frac{9(1-r^2)^3}{2(v_{\rm r}-r^3)^2} - 6\kappa_{\rm p}C_0R_0\frac{(1-r^2)^2\phi_{\rm cy}}{v_{\rm r}-r^3}  \nonumber \\
&& + 4(1-r^2)R_0^2\Big\{(\sigma_{\rm p}\phi_{\rm cy} + b \phi_{\rm cy}^2) \\ \nonumber
&&+  \frac{ k_{\rm B}T}{a_{\rm p}}[\phi_{\rm cy} \ln(\phi_{\rm cy}) +  (1-\phi_{\rm cy}) \ln(1-\phi_{\rm cy}) ]\Big\},
\end{eqnarray} 
where $\kappa_{\rm dif}=\kappa_{\rm p}-\kappa_{\rm d}$ and 
 $\sigma_{\rm p}= - \mu/a_{\rm p} + \kappa_{\rm p}C_0^2/2$.
The chemical potential of the protein binding is $\mu$,
and $a_{\rm p}$ is the membrane area bound by one protein.
The last terms in Eqs.~(\ref{eq:FSP0})--(\ref{eq:FCY0}) represent the mixing entropy of the bound proteins.
The inter-protein interactions are taken into account as squared density terms of $b\phi^2$.
Proteins have repulsive or attractive interactions at $b>0$ and $b<0$, respectively.

The densities $\phi_{\rm sp}$ and $\phi_{\rm cy}$ are obtained from $\partial F/\partial \phi|_{H}=0$:
\begin{equation}\label{eq:phi0}
\phi = \frac{1}{1+\exp\big[\frac{a_{\rm p}}{k_{\rm B}T}(2\kappa_{\rm dif}H^2 -2\kappa_{\rm p}C_0H + \sigma_{\rm p} + 2b\phi )\big]},
\end{equation}
where $H=1/R_{\rm sp}$ and $H=1/2R_{\rm cy}$ for the spherical and cylindrical components, respectively~\cite{nogu21a}.
For $b=0$, this is a sigmoid function of $\mu$.
For $b\ne 0$, Eq.~(\ref{eq:phi0}) is iteratively solved with an updated $\phi$ value in the right hand side.
Thus, the free energy $F$ is expressed as a function of one variable $r$ using Eqs.~(\ref{eq:FSP0}) and (\ref{eq:FCY0})
with $\phi_{\rm sp}$ and $\phi_{\rm sp}$ obtained using Eq.~(\ref{eq:phi0}).
Hence, the free-energy minimum is calculated by $\partial F/\partial r=0$.

The maximum binding (sensing) of the proteins occurs at a higher density than the curvature generation~\cite{nogu21a}.
The curvatures of sensing and generation are obtained by $\partial\phi/\partial H=0$ and  $\partial F/\partial H=0$ 
in the absence of any constraints and external forces, respectively; 
The former and latter are the  preferred curvatures for protein binding and for the entire membrane including bare membrane parts, respectively.
The sensing curvature is $H_{\rm s} = \kappa_{\rm p}C_0/2\kappa_{\rm dif}$,
and the curvature of the generation  is  $H_{\rm g} = \kappa_{\rm p}\phi C_0/2(\kappa_{\rm dif}\phi + \kappa_{\rm d})$.
Hence, the maximum protein density $\phi_{\rm cy}^{\rm max}$ is obtained at $1/R_{\rm cy} = C_{\rm s} =  \kappa_{\rm p}C_0/\kappa_{\rm dif}$ 
for the cylindrical tube:
 $\phi_{\rm cy}^{\rm max}=1/\{1+\exp[-(\mu -\mu_0 )/k_{\rm B}T]\}$ at $b=0$,
where $\mu_0 = - a_{\rm p}\kappa_{\rm p}{C_0}^2(\kappa_{\rm p}/\kappa_{\rm dif}  - 1)/2$.
At $\mu=\mu_0$, $\phi_{\rm cy}^{\rm max}=1/2$ is obtained.
In the dilute limit $\phi_{\rm sp}\ll 1$ and $\phi_{\rm sp}\ll 1$ with $R_{\rm sp} \gg R_{\rm cy}$ and $b=0$, 
the density ratio is given by
\begin{equation}
\frac{\phi_{\rm cy}}{\phi_{\rm sp}} \simeq \exp\Big\{ -\frac{\kappa_{\rm dif}a_{\rm p}}{2k_{\rm B}T}\Big[\Big(\frac{1}{R_{\rm cy}} - C_{\rm s}\Big)^2 - C_{\rm s}^2 \Big]\Big\},
\end{equation}
as reported in Ref.~\citenum{prev15}.

The surface tension $\sigma_{\rm d}$ and osmotic pressure $\Pi$ can be expressed as Lagrange multipliers to maintain the area and volume, respectively: $\breve{F}=F + \sigma_{\rm d}A - \Pi V$.
Then, $\sigma_{\rm d}$, $\Pi$, and $f_{\rm ex}$ satisfy $\partial\breve{F}/\partial R_{\rm sp}|_{R_{\rm cy},L_{\rm cy}}=0$, $\partial\breve{F}/\partial R_{\rm cy}|_{R_{\rm sp},L_{\rm cy}}=0$, and $\partial\breve{F}/\partial L_{\rm cy}|_{R_{\rm sp},R_{\rm cy}}=0$:
\begin{eqnarray}\label{eq:pres0}
\Pi &=&   \frac{f_{\rm ex}}{\pi R_{\rm cy}^2}
- \frac{2(\kappa_{\rm dif}\phi_{\rm cy}+\kappa_{\rm d})}{R_{\rm cy}^3} + \frac{2\kappa_{\rm p}C_0\phi_{\rm cy}}{R_{\rm cy}^2}, \\ \label{eq:ten0}
\sigma_{\rm d}  &=& \Pi R_{\rm cy} + \frac{\kappa_{\rm dif}\phi_{\rm cy}+\kappa_{\rm d}}{2R_{\rm cy}^2} - \sigma_{\rm p}\phi_{\rm cy} - b \phi_{\rm cy}^2 \\ \nonumber
&&-  \frac{k_{\rm B}T}{a_{\rm p}}[\phi_{\rm cy} \ln(\phi_{\rm cy}) +  (1-\phi_{\rm cy}) \ln(1-\phi_{\rm cy}) ]  \\ \label{eq:ten1}
  &=& \frac{\Pi R_{\rm sp}}{2} + \frac{f_{\rm ex}}{4\pi R_{\rm sp}} + \frac{\kappa_{\rm p}C_0\phi_{\rm sp}}{R_{\rm sp}} - \sigma_{\rm p}\phi_{\rm sp} - b \phi_{\rm sp}^2  \\ \nonumber
&&-  \frac{k_{\rm B}T}{a_{\rm p}}[\phi_{\rm sp} \ln(\phi_{\rm sp}) +  (1-\phi_{\rm sp}) \ln(1-\phi_{\rm sp}) ].
\end{eqnarray} 
The first terms in Eqs.~(\ref{eq:ten0}) and (\ref{eq:ten1}) represent the Laplace tensions.
Here, the spherical and cylindrical components share the same surface tension, which is different from the analysis in Ref.~\citenum{prev15}.
Because lipid molecules can freely move between these components, their surface tension is balanced.
Note that proteins can freely move between these components through binding and unbinding with bulk diffusion in addition to the surface diffusion.

\begin{figure}
\includegraphics[]{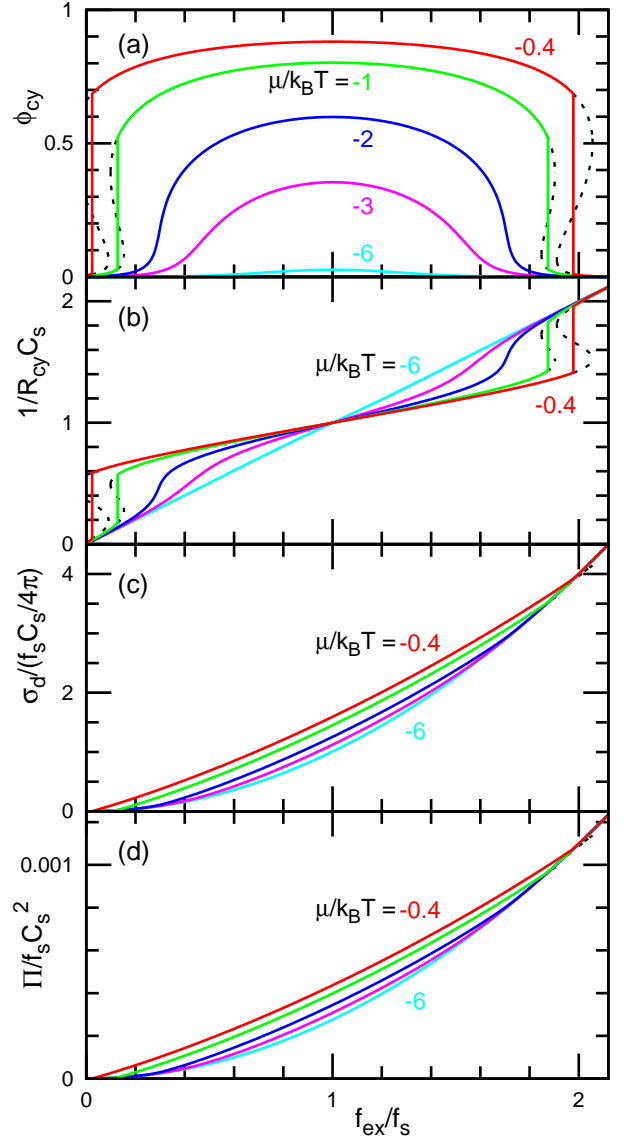}
\caption{
Force $f_{\rm ex}$ dependence of (a) the protein density $\phi_{\rm cy}$ and (b) curvature $1/R_{\rm cy}$ of the cylindrical membrane, (c) the surface tension $\sigma_{\rm d}$, and (d) the osmotic pressure $\Pi$
for $\mu/k_{\rm B}T=-6$, $-3$, $-2$, $-1$, and $-0.4$ at $C_0R_0=400$, $v_{\rm r}=0.9$, $\kappa_{\rm p}/\kappa_{\rm d}=3$, and $b=0$.
The solid lines represent thermal equilibrium states.
The dashed lines represent the metastable and free-energy-barrier states.
From the bottom to top, $\mu/k_{\rm B}T=-6$, $-3$, $-2$, $-1$, and $-0.4$ in (a), (c), and (d).
}
\label{fig:v9c4f}
\end{figure}

\begin{figure}
\includegraphics[]{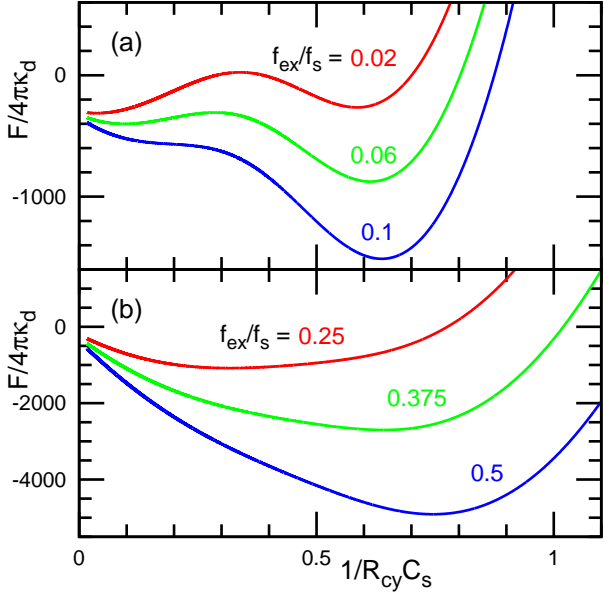}
\caption{
Free energy profile at $C_0R_0=400$, $v_{\rm r}=0.9$, $\kappa_{\rm p}/\kappa_{\rm d}=3$, and $b=0$.
(a) $\mu/k_{\rm B}T=-0.4$. (b) $\mu/k_{\rm B}T=-2$.
The free energy has a double minima at $f_{\rm ex}/f_{\rm s} = 0.02$ and $0.06$ in (a).
}
\label{fig:en}
\end{figure}

 When the volume of the cylindrical tube is
negligibly small as $R_{\rm cy}^2L_{\rm cy}/R_0^3 \ll 1$,
the spherical component can be approximated to have a maximum volume of $V=4\pi {R_{\rm sp}}^3/3$.
In this limit condition~\cite{smit04}, 
the following relation is obtained:
\begin{eqnarray}\label{eq:vt}
  r &\simeq& v_{\rm r}^{1/3}, \\
  \frac{R_{\rm cy}L_{\rm cy}}{R_0^2} &\simeq&  2(1- v_{\rm r}^{2/3}).
\end{eqnarray} 
Smith et al. reported that this approximation reproduces the force-length relation 
of tethered vesicles with zero spontaneous curvature~\cite{smit04}.
In this approximation, $L_{\rm cy}$ is a variable, whereas $r$ is a constant.
We call this approximation limit analysis and use sub- or superscripts lim to represent the quantities calculated by this method.
The vesicle shape is obtained from $\partial F/\partial L_{\rm cy}=0$, as
\begin{eqnarray}
f_{\rm ex} &\simeq& \frac{\pi(\kappa_{\rm dif}\phi_{\rm cy}+\kappa_{\rm d})L_{\rm cy}}{(1 - v_{\rm r}^{2/3})R_0^2}
- 2\pi\kappa_{\rm p}C_0\phi_{\rm cy}  \label{eq:fext1} \\ \label{eq:fext2}
 &=& \frac{2\pi(\kappa_{\rm dif}\phi_{\rm cy}+\kappa_{\rm d})}{R_{\rm cy}}
- 2\pi\kappa_{\rm p}C_0\phi_{\rm cy}  \\ \label{eq:fext3}
 &=&  \Big[ \Big(\frac{\kappa_{\rm dif}}{\kappa_{\rm d}}\phi_{\rm cy}+1 \Big)\Big(\frac{1}{R_{\rm cy}C_{\rm s}}  - 1\Big) +1 \Big] f_{\rm s},
\end{eqnarray}
where 
\begin{equation}
f_{\rm s} = 2\pi \kappa_{\rm d}C_{\rm s}.
\end{equation}
The maximum protein density is obtained at $f_{\rm ex} = f_{\rm s}$, since $1/R_{\rm cy }=C_{\rm s}$ at the maximum.
The force $f_{\rm ex}$  linearly increases with the tube curvature $1/R_{\rm cy}$ for a constant $\phi_{\rm cy}$.
In particular, for completely unbound tubes ($\phi_{\rm cy}=0$),
$f_{\rm ex}$ is proportional to $1/R_{\rm cy}$  as $f_{\rm ex}/f_{\rm s}  \simeq 1/R_{\rm cy}C_{\rm s}$.
For any value of $\phi_{\rm cy}$,  $f_{\rm ex}=f_{\rm s}$ at $R_{\rm cy}C_{\rm s}=1$.
The surface tension  $\sigma_{\rm d}$ is obtained from $\partial \breve{F}/\partial R_{\rm cy}|_{L_{\rm cy}}=0$
with $\breve{F}=F + \sigma_{\rm d}A$:
\begin{eqnarray} \label{eq:tent}
\sigma_{\rm d} &\simeq& \frac{\kappa_{\rm dif}\phi_{\rm cy}+\kappa_{\rm d}}{2R_{\rm cy}^2} 
-  \sigma_{\rm p}\phi_{\rm cy} - b \phi_{\rm cy}^2 \\ \nonumber
&& -  \frac{k_{\rm B}T}{a_{\rm p}}[\phi_{\rm cy} \ln(\phi_{\rm cy}) +  (1-\phi_{\rm cy}) \ln(1-\phi_{\rm cy}) ]. 
\end{eqnarray}
Hence, the first term in Eq.~(\ref{eq:ten0}) is neglected in the limit analysis.
For completely unbound membrane tubes of $\phi_{\rm cy}=0$, a well-known relation
$\sigma_{\rm d}=\kappa_{\rm d}/2R_{\rm cy}^2 = f_{\rm ex}/4\pi R_{\rm cy}$  is obtained.
This relation has been used to estimate $R_{\rm cy}$ and $\kappa_{\rm d}$ from $\sigma_{\rm d}$ and $f_{\rm ex}$ experimentally~\cite{dimo14,bo89,evan96,cuve05}.
However, the surface tension of the bound membranes exhibits a more complicated dependence, as expressed in Eq.~(\ref{eq:tent}).
As described in Sec.~\ref{sec:tres}, this limit method provides a good  approximation for thin tubes.

We use  $\kappa_{\rm d} = 20 k_{\rm B}T = 8 \times 10^{-20}$\,J, $a_{\rm p} = 100$\,nm$^2$, and $R_0= 10\,\mu$m,
i.e., $a_{\rm p}/R_0^2= 10^{-6}$.
External forces are typically at $f_{\rm ex} \sim 10$\,pN, which can be imposed using optical tweezers.
For $C_{\rm s}= 100/R_0 = 10^{7} {\rm m}^{-1}$,
$f_{\rm s}=2\pi \kappa_{\rm d}C_{\rm s} \simeq 5$\,pN.
In the plots, the data at $R_{\rm cy}/R_0<0.1$ are displayed (i.e., too wide tubes are excluded).
In this study, the iterations of the density calculation for $b\ne 0$ were repeated until the difference in $\phi$ was less than $10^{-8}$;
fewer than ten iterations were typically performed.

\begin{figure}
\includegraphics[]{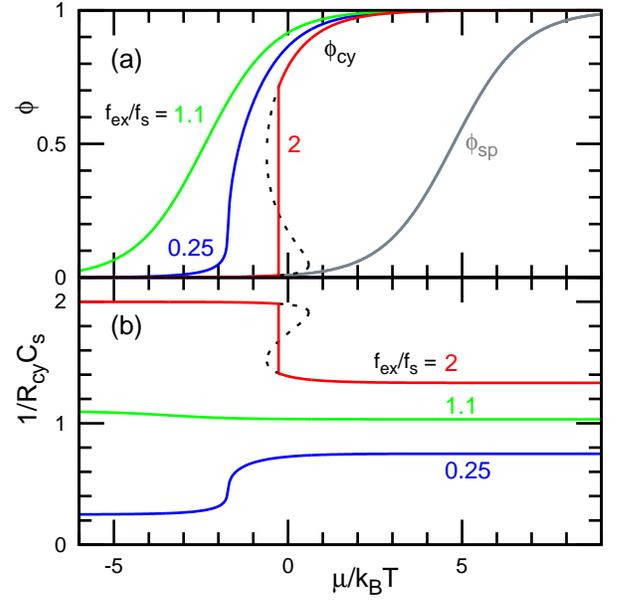}
\caption{
Chemical potential  $\mu$ dependence of (a) the protein densities $\phi$ and (b) the curvature $1/R_{\rm cy}$ of the cylindrical membrane for $f_{\rm ex}/f_{\rm s} = 0.25$, $1.1$, and $2$ at $C_0R_0=400$, $v_{\rm r}=0.9$, $\kappa_{\rm p}/\kappa_{\rm d}=3$, and $b=0$.
The solid lines represent thermal equilibrium states.
The dashed lines represent the metastable and free-energy-barrier states.
The gray line in (a) represents $\phi_{\rm sp}$ (all data for $f_{\rm ex}/f_{\rm s} = 0.25$, $1.1$, and $2$ 
overlap this single curve).
}
\label{fig:v9c4m}
\end{figure}

\begin{figure}
\includegraphics[]{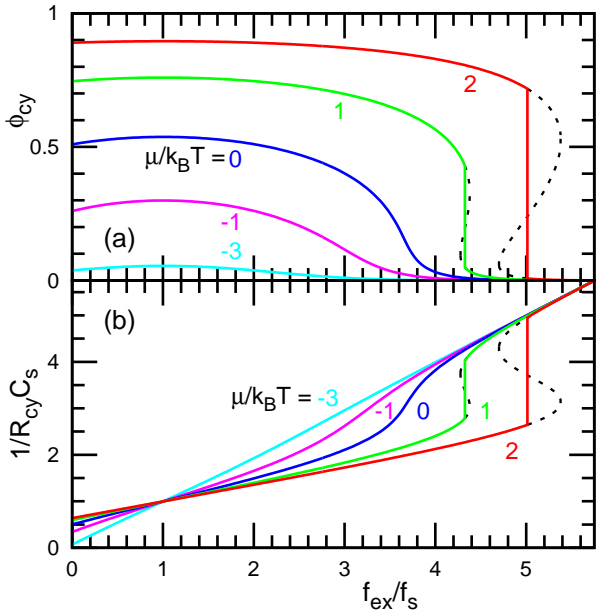}
\caption{
Force $f_{\rm ex}$ dependence of (a) the protein density $\phi_{\rm cy}$ and (b) curvature $1/R_{\rm cy}$ of the cylindrical membrane
for $\mu/k_{\rm B}T=-3$, $-1$, $0$, $1$, and $2$ at $C_0R_0=100$, $v_{\rm r}=0.9$, $\kappa_{\rm p}/\kappa_{\rm d}=3$, and $b=0$.
The solid lines represent thermal equilibrium states.
The dashed lines represent the metastable and free-energy-barrier states.
}
\label{fig:v9c1f}
\end{figure}

\begin{figure}
\includegraphics[]{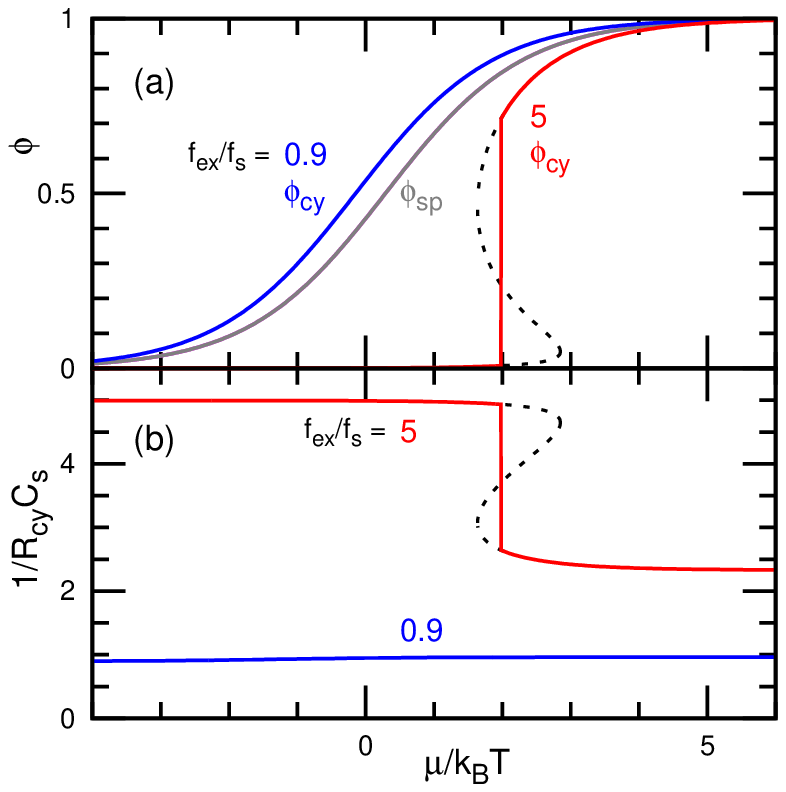}
\caption{
Chemical potential  $\mu$ dependence of (a) the protein densities $\phi$ and (b) the curvature $1/R_{\rm cy}$ of the cylindrical membrane for $f_{\rm ex}/f_{\rm s} = 0.9$ and $5$ at $C_0R_0=100$, $v_{\rm r}=0.9$, $\kappa_{\rm p}/\kappa_{\rm d}=3$, and $b=0$.
The solid lines represent thermal equilibrium states.
The dashed lines represent the metastable and free-energy-barrier states.
The gray line in (a) represents $\phi_{\rm sp}$ (both data for $f_{\rm ex}/f_{\rm s} = 0.9$ and $5$ 
overlap this single curve).
}
\label{fig:v9c1m}
\end{figure}

\begin{figure}
\includegraphics[]{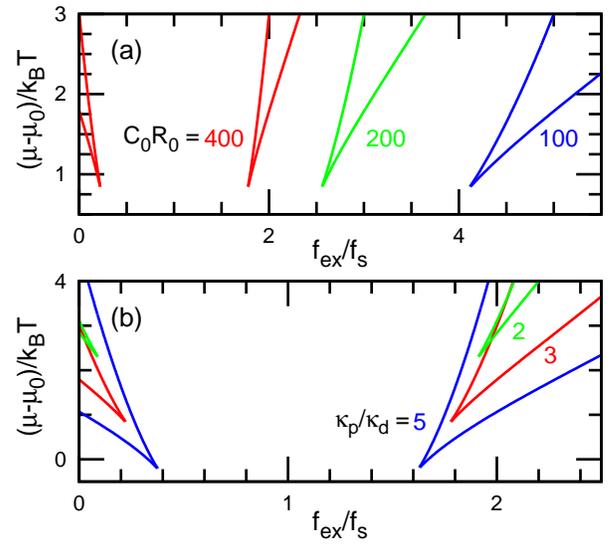}
\caption{
Phase diagrams at $v_{\rm r}=0.9$ and $b=0$.
Two states with narrow and wide tubes coexist in
the region between the two solid lines.
(a) $C_0R_0=100$, $200$, and $400$ at  $\kappa_{\rm p}/\kappa_{\rm d}=3$.
(b) $\kappa_{\rm p}/\kappa_{\rm d}=2$, $3$, and $5$ at $C_0R_0=400$.
The origin of the chemical potential is at $\mu_0$ to give $\phi_{\rm cy}=0.5$ at $f_{\rm ex}/f_{\rm s}=1$.
}
\label{fig:pd}
\end{figure}

\begin{figure}
\includegraphics[]{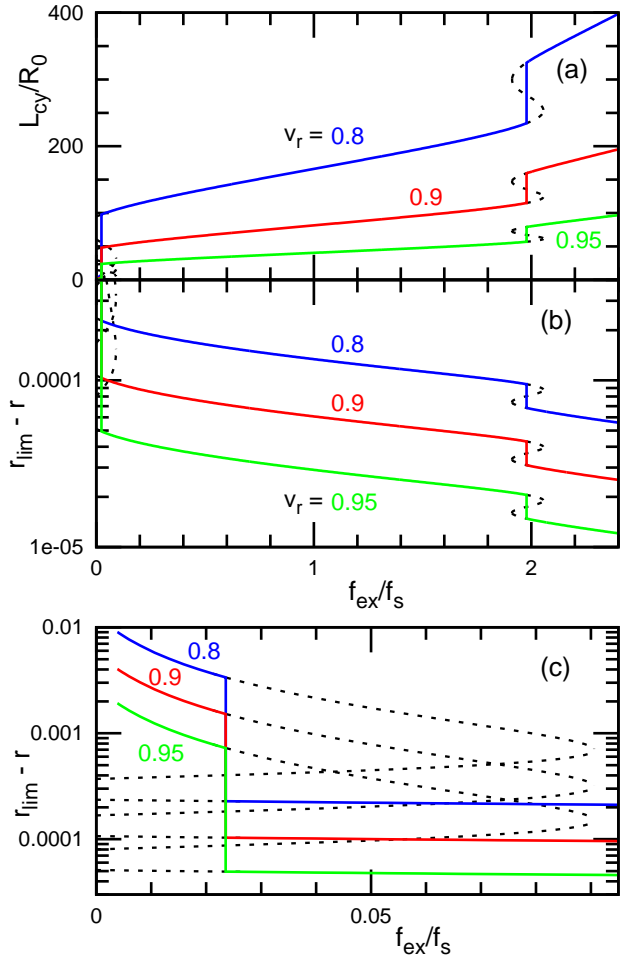}
\caption{
Effects of reduced volume $v_{\rm r}$ at  $C_0R_0=400$, $\mu/k_{\rm B}T=-0.4$, $\kappa_{\rm p}/\kappa_{\rm d}=3$, and $b=0$.
(a) Length of the cylindrical membrane for $v_{\rm r}=0.8$, $0.9$, and $0.95$.
(b),(c) Difference of radius of the spherical membrane from the maximum value $r_{\rm lim}$
for (b) large and (c) small $f_{\rm ex}$ regions.
The solid lines represent thermal equilibrium states.
The dashed lines represent the metastable and free-energy-barrier states.
}
\label{fig:v}
\end{figure}

\begin{figure}
\includegraphics[]{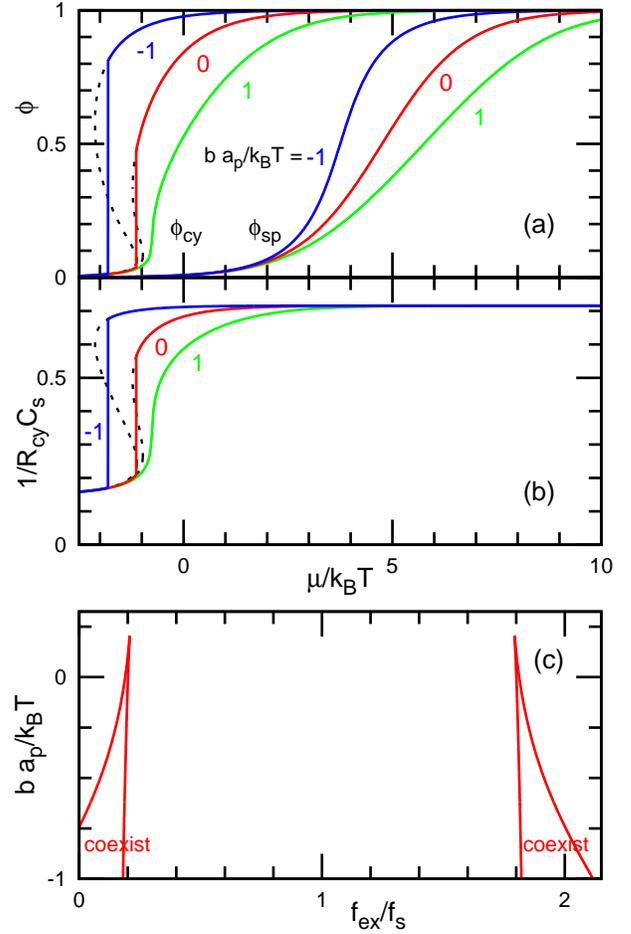}
\caption{
Dependence on pairwise interactions at $C_0R_0=400$, $v_{\rm r}=0.9$, and $\kappa_{\rm p}/\kappa_{\rm d}=3$.
(a) Protein densities $\phi_{\rm cy}$ and $\phi_{\rm sp}$ of the cylindrical and spherical membrane components, respectively,
for  $b a_{\rm p}/k_{\rm B}T= -1$, $0$, and $1$ at $f_{\rm ex}/f_{\rm s}=0.15$.
(b) The curvature $1/R_{\rm cy}$ of cylindrical membrane for  $b a_{\rm p}/k_{\rm B}T= -1$, $0$, and $1$ at $f_{\rm ex}/f_{\rm s}=0.15$.
(c) Phase diagrams at $\mu/k_{\rm B}T=-1.5$.
Two states with narrow and wide tubes coexist in
the region between the two solid lines.
}
\label{fig:b0}
\end{figure}

\begin{figure}
\includegraphics[]{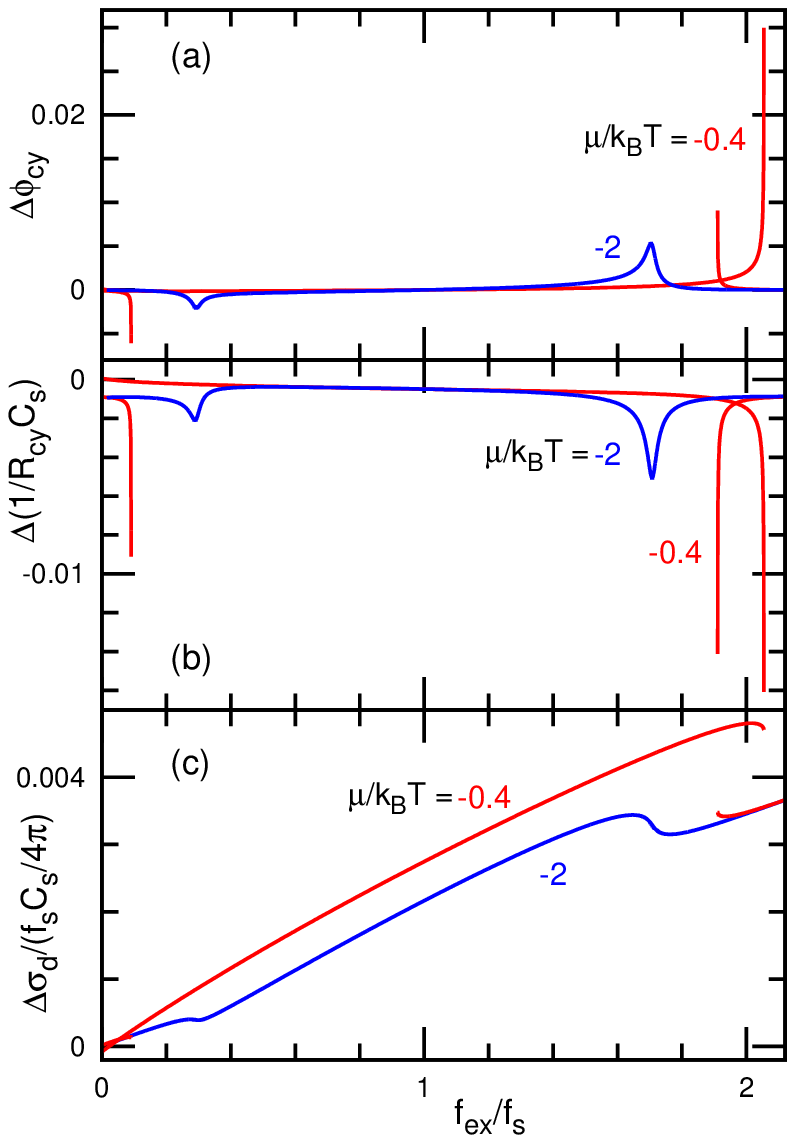}
\caption{
Accuracy of the limit analysis using Eqs.~(\ref{eq:vt})--(\ref{eq:tent}) for $\mu/k_{\rm B}T=-2$ and $-0.4$  at $C_0R_0=400$, $v_{\rm r}=0.9$, $\kappa_{\rm p}/\kappa_{\rm d}=3$, and $b=0$.
(a) Difference of protein density $\Delta \phi_{\rm cy} = \phi_{\rm cy} - \phi_{\rm cy}^{\rm lim}$.
(b) Difference of the curvature of cylindrical membrane $\Delta (1/R_{\rm cy}C_{\rm s}) = 1/R_{\rm cy}C_{\rm s} - 1/R_{\rm cy}^{\rm lim}C_{\rm s}$.
(c) Difference of the surface tension $\Delta \sigma_{\rm d} = \sigma_{\rm d} - \sigma_{\rm d}^{\rm lim}$.
The superscript lim represents data obtained by the limit analysis.
The thermal equilibrium and metastable states are used for $\mu/k_{\rm B}T=-0.4$.
}
\label{fig:v9c4t}
\end{figure}

\subsection{Theoretical results}\label{sec:tres}

In this subsection, we describe the results of the mean-field theory using
the simplified geometry depicted in Fig.~\ref{fig:cat}.
Figures~\ref{fig:v9c4f}--\ref{fig:v9c4m} 
and Figs.~\ref{fig:v9c1f}--\ref{fig:v9c1m} show the results
at $C_0R_0=400$ and  $C_0R_0=100$, respectively, at $v_{\rm r}=0.9$, $\kappa_{\rm p}/\kappa_{\rm d}=3$, and $b=0$.
When the proteins are unbound at a low chemical potential $\mu$, 
the tube curvature is proportional to the force strength,
 as shown in Figs.~\ref{fig:v9c4f}(b) and \ref{fig:v9c1f}(b).
As  $\mu$ increases, the protein density $\phi_{\rm cy}$   increases, and
the force curve  changes, as  predicted in Eqs.~(\ref{eq:phi0}) and (\ref{eq:fext3}).
The force dependence curves of $\phi_{\rm cy}$ are visually reflection symmetric with respect to $f_{\rm ex}/f_{\rm s}=1$
and take the maxima at $f_{\rm ex}/f_{\rm s}=1$  as shown in Figs.~\ref{fig:v9c4f}(a) and \ref{fig:v9c1f}(a).
The tube curvature $1/R_{\rm cy}$ is point symmetric with respect to $f_{\rm ex}/f_{\rm s}=1$ (see Figs.~\ref{fig:v9c4f}(b) and \ref{fig:v9c1f}(b)).
Thus, the results are reproduced well by the limit analysis using $r=v_{\rm r}^{1/3}$.
When $C_0$ and $\kappa_{\rm p}$ are unknown, they can be estimated from the tube radius at the maximum value of $\phi_{\rm cy}$
and the slope of $d(1/R_{\rm cy})/df_{\rm ex} = 1/2\pi(k_{\rm dif}\phi_{\rm cy}^{\rm max}+k_{\rm d})$ at $f_{\rm ex}/f_{\rm s}=1$, respectively.

For large  $\mu$, first-order transitions occur between the unbound and bound tubes with different tube radii
(see Figs.~\ref{fig:v9c4f} and \ref{fig:v9c1f}).
These transitions appear symmetrically with both small and large $f_{\rm ex}$ with respect to $f_{\rm ex}/f_{\rm s}=1$.
Around the transition points, the free energy profile has two minima (see Fig.~\ref{fig:en}), 
so that the unbound and bound states coexist (see 
 van der Waals loop depicted by the dashed lines in Figs.~\ref{fig:v9c4f} and \ref{fig:v9c1f}).
These two transitions can be understood using Eq.~(\ref{eq:fext3}) of the limit analysis;
at large $\mu$, Eq.~(\ref{eq:fext3}) has two regions of $df_{\rm ex}/d(1/R_{\rm cy})<0$, which
means an unstable solution, i.e., free-energy barrier states between stable and metastable phases.
With increasing $\mu$, the  $f_{\rm ex}$ width of the coexistence increases, and
the range of $f_{\rm ex}$ is shifted outward (see Fig.~\ref{fig:pd}).
At small spontaneous curvatures ($C_0R_0 \lesssim 200$), the transition at a small  $f_{\rm ex}$ disappears,
since it moves into the unphysical region ($f_{\rm ex}<0$).
Although the transition at $f_{\rm ex}< f_{\rm s}$ has been previously reported in Ref.~\citenum{prev15},
the reentrant transition  at $f_{\rm ex}> f_{\rm s}$ has not yet been reported.
These transitions are similar to the budding transition between a small number of large buds and a large number of small buds~\cite{nogu21a}.
However, one of the specific features of the present case is that the transitions occur twice with increasing force.
This is because tubes with larger curvatures than $C_{\rm s}$ are generated by large external forces, unlike spontaneous budding.

The density $\phi_{\rm sp}$ of the spherical component follows the sigmoid function $1/\{1+\exp[-(\mu-\mu_{\rm half})/k_{\rm B}T]\}$,
as shown in Figs.~\ref{fig:v9c4m} and \ref{fig:v9c1m}.
Since changes in the radius $R_{\rm sp}$ are very small (see Figs.~\ref{fig:v}(b) and (c)), 
the deviation from this sigmoid function is negligibly small.
In contrast,  $\phi_{\rm cy}$--$\mu$ curves can largely deviate from the  sigmoid function accompanied by changes in $R_{\rm cy}$ (see Figs.~\ref{fig:v9c4m} and \ref{fig:v9c1m}).
Interestingly, the proteins can bind more onto the spherical component than onto the cylindrical tube
as shown in Fig.~\ref{fig:v9c1m}(a). This occurs in tubes that are narrower than $1/C_{\rm s}$ at large $f_{\rm ex}$ values.

With increasing force $f_{\rm ex}$, the surface tension $\sigma_{\rm d}$ and osmotic pressure $\Pi$ increase together.
For $\phi_{\rm cy} \simeq 0$ at a low $\mu$, $\sigma_{\rm d}= f_{\rm ex}/4\pi R_{\rm cy}$ is obtained.
At larger $\phi_{\rm cy}$ (due to large $\mu$), $\sigma_{\rm d}$ and $\Pi$ slightly increase (see Figs.~\ref{fig:v9c4f}(c) and (d)).
The vesicle ruptures when $\sigma_{\rm d}$ overcomes the lysis tension, which is typically $1$--$25$\, mN/m,
 depending on the membrane composition and conditions~\cite{evan00,evan03,ly04}.
Since the maximum value in Fig.~\ref{fig:v9c4f}(c) is $\sigma_{\rm d} \simeq 0.08$\,mN/m,
the lipid membranes are not yet ruptured in this range.
However, the tubular membranes may become unstable.
The minimum radius of the tubular membranes is $\sim 10$\,nm,  depending on the membrane composition.
In the right end regions in Figs.~\ref{fig:v9c4f} and \ref{fig:v9c1f},
the tubular membrane can be ruptured or the tubular radius is saturated to a finite value due to the repulsion between membranes:
For $C_0R_0=400$, $R_{\rm cy}= 1/2C_{\rm s} \simeq 8$\,nm and 
for  $C_0R_0=100$, $R_{\rm cy}= 1/5C_{\rm s} \simeq 13$\,nm.

As the bending rigidity $\kappa_{\rm p}$ of the bound membrane increases,
the slope of $1/R_{\rm cy}$--$f_{\rm ex}$ curves at $\phi_{\rm cy}=1$ decreases, and 
the first-order transitions start at a lower $\phi_{\rm cy}^{\rm max}$ (see Fig.~\ref{fig:pd}(b)).
As the reduced volume $v_{\rm r}$ decreases,
the tube length $L_{\rm cy}$ increases owing to the larger available area of the tube (see Fig.~\ref{fig:v}(a)).
However, the tube radius $R_{\rm cy}$ remains almost unchanged.

Next, we consider the inter-protein interactions ($b \ne 0$).
As $b$ decreases, protein binding is promoted at large $\phi$ (Figs.~\ref{fig:b0}(a),(b)),
such that the coexistence region widens (Fig.~\ref{fig:b0}(c)).
The reflection and point symmetries of the $\phi_{\rm cy}$--$f_{\rm ex}$ and  $1/R_{\rm cy}$--$f_{\rm ex}$  curves (Figs.~\ref{fig:v9c4f}(a) and (b))
remain unchanged for $b \ne 0$, respectively (data not shown).
Note that the membrane exhibits a phase separation with large and small $\phi$ within each component at $b < -2 k_{\rm B}T/a_{\rm p}$, 
in which $F_{\rm sp}$ and $F_{\rm cy}$ can have double minima~\cite{nogu21a}. Thus, we only consider $b > -2 k_{\rm B}T/a_{\rm p}$ in this study.

To confirm the quality of the limit analysis,
the deviations of $\phi_{\rm cy}$, $1/R_{\rm cy}$, and $\sigma_{\rm d}$ are shown in Fig.~\ref{fig:v9c4t}.
All of these are very small. Note that the larger values at the ends of the curves in Figs.~\ref{fig:v9c4t}(a) and (b)
are due to slight shifts in the spinodal points along $f_{\rm ex}$.
Thus, this method provides an accurate approximation, so that one can focus only on the cylindrical component
to study the protein binding on the tethered vesicle.
In the next section, we simulate only the membrane tubes.

\section{Simulation of membrane tubes}\label{sec:sim}

\subsection{Simulation model}

A fluid membrane is represented by a self-assembled single-layer sheet of $N$ particles.
The position and orientational vectors of the $i$-th particle are ${\bm{r}}_{i}$ and ${\bm{u}}_i$, respectively.
The details of the spin meshless membrane model are described in Ref.~\citenum{shib11},
and the combination with the protein binding is described in Ref.~\citenum{gout21};
the model is described briefly here.

\begin{figure*}
\includegraphics[]{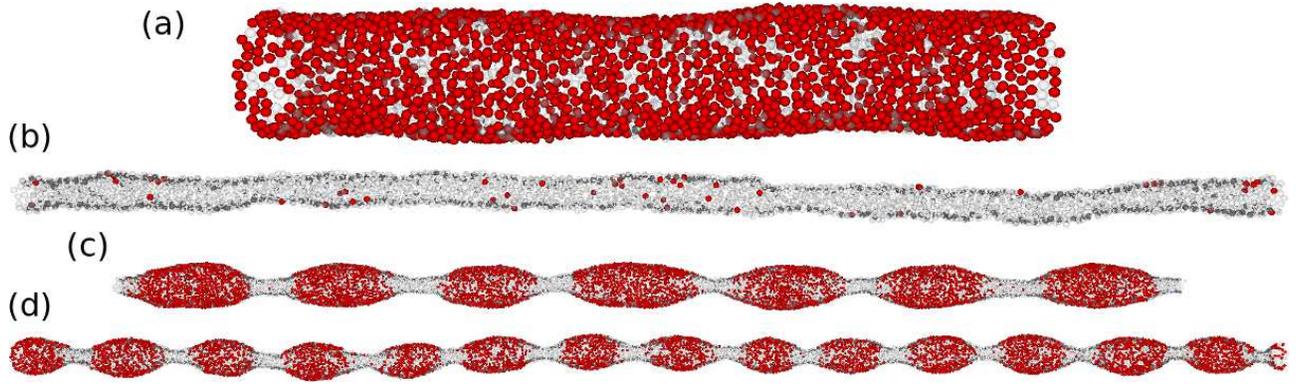}
\caption{
  Snapshots of the meshless membrane simulations at (a),(b) $N=2400$ and (c),(d) $N=9600$.  
(a),(b) At $f_{\rm ex}\sigma/k_{\rm B}T=43$, $\mu_{\rm ms}/k_{\rm B}T=3.5$, and $C_0\sigma=0.05$,
(a) a wide tube with a high protein density and (b) narrow tube with a low protein density
coexist.
(c),(d) Phase-separated membranes for (c) $C_0\sigma=0.05$
and (d) $C_0\sigma=0.075$ at $f_{\rm ex}\sigma/k_{\rm B}T=38$ and $\mu_{\rm ms}/k_{\rm B}T=3$.
Red and transparent gray spheres represent the bound and unbound membrane particles, 
respectively. 
}
\label{fig:snaps}
\end{figure*}

\begin{figure}
\includegraphics[]{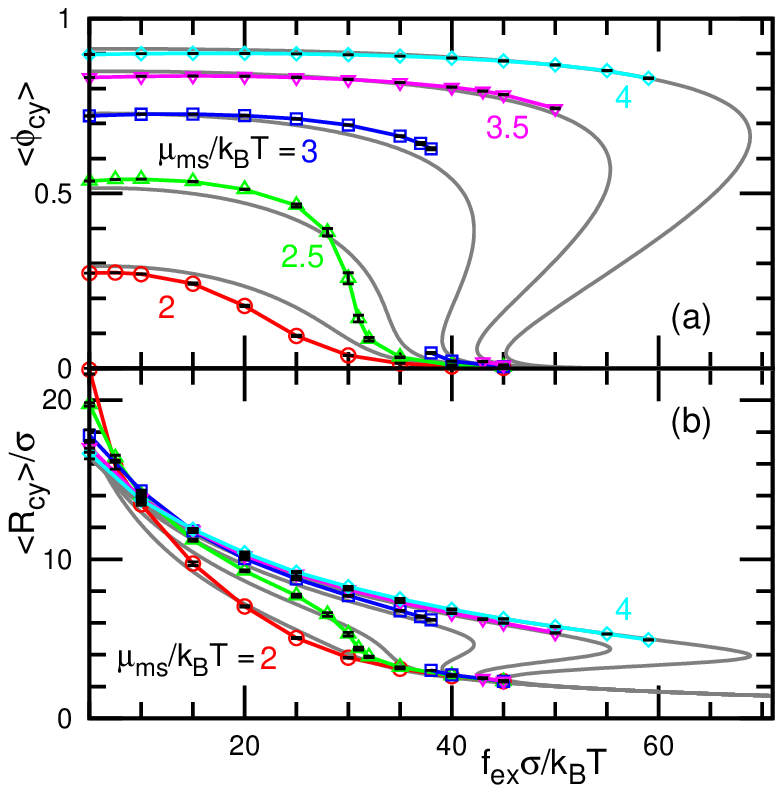}
\caption{
Tubular membranes for $\mu_{\rm ms}/k_{\rm B}T=2$, $2.5$, $3$, $3.5$, and $4$  at $C_0\sigma=0.05$.
(a) Protein density $\phi_{\rm cy}$. (b) Tube radius $R_{\rm cy}$. 
The symbols with solid lines represent simulation data,
and the gray lines represent the prediction of the mean-field theory
at $\mu= \mu_{\rm ms} -3.5k_{\rm B}T$ and $b a_{\rm p}/k_{\rm B}T=-1$.
}
\label{fig:tubec1}
\end{figure}

\begin{figure}
\includegraphics[]{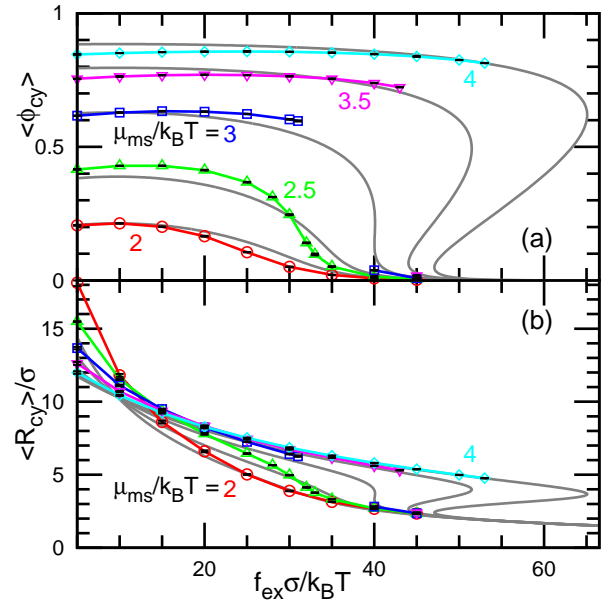}
\caption{
Tubular membranes for $\mu_{\rm ms}/k_{\rm B}T=2$, $2.5$, $3$, $3.5$, and $4$  at $C_0\sigma=0.075$.
(a) Protein density $\phi_{\rm cy}$. (b) Tube radius $R_{\rm cy}$. 
The symbols with solid lines represent simulation data,
and the gray lines represent the prediction of the mean-field theory
at $\mu= \mu_{\rm ms} -3.8k_{\rm B}T$ and $b a_{\rm p}/k_{\rm B}T=-1$.
}
\label{fig:tubec15}
\end{figure}

The membrane particles interact with each other via the potential $U=U_{\rm rep}+U_{\rm att}+U_{\rm bend}+U_{\rm tilt}$.
The potential $U_{\rm rep}$ is an excluded volume interaction with diameter $\sigma$ for all pairs of particles.
The solvent is implicitly accounted for by the effective attractive potential $U_{\rm att}$.
The bending and tilt potentials
are given as follows: 
\begin{eqnarray}
\frac{U_{\rm bend}}{k_{\rm B}T} &=& \frac{k_{\rm bend}}{2} \sum_{i<j} ({\bm{u}}_{i} - {\bm{u}}_{j} - C_{\rm bd} \hat{\bm{r}}_{i,j} )^2 w_{\rm cv}(r_{i,j}), \\
\frac{U_{\rm tilt}}{k_{\rm B}T} &=& \frac{k_{\rm tilt}}{2} \sum_{i<j} [ ( {\bm{u}}_{i}\cdot \hat{\bm{r}}_{i,j})^2
 + ({\bm{u}}_{j}\cdot \hat{\bm{r}}_{i,j})^2  ] w_{\rm cv}(r_{i,j}),\hspace{0.7cm} 
\end{eqnarray}
  respectively, where
 $\hat{\bm{r}}_{i,j}={\bm{r}}_{i,j}/r_{i,j}$, ${\bm{r}}_{i,j}= {\bm{r}}_{i}-{\bm{r}}_{j}$, and $w_{\rm cv}(r_{i,j})$ is a weight function. 
The spontaneous curvature is given by $C_0 = C_{\rm bd}/2 \sigma$~\cite{shib11}. 

Each membrane particle is a binding site and takes two states (bound and unbound).
In this study, $C_0=0$ and $k_{\rm bend}=k_{\rm tilt}=10$ for the unbound membrane particles
and $k_{\rm  bend}=k_{\rm tilt}=40$ for the bound membrane particles, where
 $\kappa_{\rm d}/k_{\rm B}T=16 \pm 1$ and $\kappa_{\rm p}/k_{\rm B}T=71 \pm 3$.
In the bending and tilt potentials, for a pair of neighboring bound and unbound particles,
we use the mean value {$k_{\rm bend}=k_{\rm tilt}=25$}.
For the bound membrane, $C_0\sigma=0.05$ and $0.075$ are used.
The ratio of the Gaussian modulus $\bar{\kappa}$ to $\kappa$ is constant as follows: 
$\bar{\kappa}/\kappa=-0.9\pm 0.1$~\cite{nogu19}.
The other parameters are the same as those used in Ref.~\citenum{gout21}.
For the unbound and bound particles, the membrane areas per particle are
 $1.251\sigma^2$ and $1.262\sigma^2$, respectively, for tensionless membranes at $C_0=0$. 

A tubular membrane consisting of $N$ particles is set along the $x$-axis connected by the periodic boundary condition.
The force $f_{\rm ex}$ is imposed along the  $x$-axis, such that the tube length fluctuates thermally.
Since the solvent is not explicitly taken into account,
the volume of the membrane tube can freely change, as assumed in the limit analysis.
We mainly used $N=2400$. In addition, four-fold longer tubes with $N=9600$ were used at several parameter sets,
to examine finite-size effects. For $f_{\rm ex}\sigma/k_{\rm B}T \leq 2$,  $N=19200$ was also used.
Membrane motion is solved by molecular dynamics with a Langevin thermostat~\cite{alle87,fell95,nogu11,nogu12}.
The bound and unbound states are stochastically switched by a Metropolis Monte Carlo procedure
with $\Delta H= \Delta U - \mu_{\rm ms}$, where $\Delta U$ is the energy difference between the bound and unbound states
and $\mu_{\rm ms}$ is the binding chemical potential of the membrane particles~\cite{gout21}.
Error bars are estimated from three independent runs.

\subsection{Simulation results}\label{sec:sres}

Simulation results are shown in Figs.~\ref{fig:snaps}--\ref{fig:tubec15}.
As $f_{\rm ex}$ increases, the protein density $\phi_{\rm cy}$ and tube radius $R_{\rm cy}$ decrease.
For $\mu_{\rm ms}/k_{\rm B}T= 3$ and $3.5$ at $C_0\sigma=0.05$, two states (large $R_{\rm cy}$ with high $\phi_{\rm cy}$ and small $R_{\rm cy}$ with low $\phi_{\rm cy}$) coexist (see Figs.~\ref{fig:snaps}(a), (b), and \ref{fig:tubec1}).
These results show very good agreement with the results of the mean-field theory with the following parameter sets (see Figs.~\ref{fig:tubec1} and \ref{fig:tubec15}).
Since the binding/unbinding processes in the simulation involve the energy change in the other potentials ($U_{\rm rep}$ and $U_{\rm att}$)
via a small area change,
the chemical potential is shifted from the input value $\mu_{\rm ms}$~\cite{gout21}.
Thus, we use $\mu=\mu_{\rm ms} -3.5k_{\rm B}T$ and $\mu_{\rm ms}-3.8k_{\rm B}T$ with $a_{\rm p}=1.2\sigma^2$ for $C_0\sigma=0.05$ and $0.075$, respectively.
In the simulation, no direct interactions are considered between the bound sites.
However, the Casimir-like attractive interactions occur between them owing to the bending rigidity difference~\cite{gout21}.
To mimic this,  $b a_{\rm p}/k_{\rm B}T=-1$ is used here.
In the theory, $\phi_{\rm cy}$ has a maximum at $f_{\rm ex} = f_{\rm s}$ ($f_{\rm s}\sigma/k_{\rm B}T=6.5$ and $9.7$ for $C_0\sigma=0.05$ and $0.075$, respectively).
Indeed, $\phi_{\rm cy}^{\rm max}$ is obtained at $f_{\rm ex} \simeq f_{\rm s}$ in the simulation for $\phi_{\rm cy}^{\rm max} \lesssim 0.5$ at $\mu_{\rm ms} = 2k_{\rm B}T$;
however, it slightly increases with increasing $\mu_{\rm ms}$ to $f_{\rm ex} \simeq 2f_{\rm s}$  at $\mu_{\rm ms} = 4k_{\rm B}T$.
Because the  Casimir-like forces are not pairwise, their multibody interactions likely induce this dependency.
At a small force $f_{\rm ex}$ and low $\phi_{\rm cy}$, the radius $R_{\rm cy}$ calculated by the simulation 
is slightly larger than the theoretical prediction.
This is due to the thermal undulation of the membrane.
Except for these, the simulation results are well reproduced.

Homogeneous membrane tubes are not maintained in some ranges of the simulation parameters.
First, the membrane tubes become unstable at  large and small limits of $f_{\rm ex}$.
The membranes are ruptured for radii that are too small $R_{\rm cy}\lesssim 2\sigma$ at large $f_{\rm ex}$.
At $f_{\rm ex}\sigma/k_{\rm B}T \lesssim 1$,
the cylindrical membrane becomes unstable and divided into vesicles (see Movie 1 provided in ESI for $f_{\rm ex}=0$, $\mu_{\rm ms} = 4k_{\rm B}T$, $C_0\sigma=0.075$, and $N=19200$).
At $f_{\rm ex}=0$, a membrane tube can form an unduloid shape~\cite{kenm03,nait95},
in which a constant value of the mean curvature $H$ is maintained everywhere.
The cylindrical tube at $f_{\rm ex}=0$ has the curvature of the curvature generation, as
 $1/R_{\rm cy} = 2H_{\rm g}=\kappa_{\rm p}\langle \phi_{\rm cy}\rangle C_0/(\kappa_{\rm dif}\langle \phi_{\rm cy}\rangle + \kappa_{\rm d})$
from Eq.~(\ref{eq:fext2}).
However, longer tubes than the unduloid wavelength $l_{\rm und}=2\pi R_{\rm cy}$  (large $\mu_{\rm ms}$ at $N=9600$ and $19200$)
deform into unduloid shapes
and subsequent membrane fission at the narrow neck of the tube leads to the formation of spherical vesicles.
Cylindrical shapes maintain for  shorter tubes than the unduloid wavelength
owing to the finite-size effect.

In addition, phase separation can unstabilize cylindrical shapes.
A beaded-necklace-like tube with phase separation is formed in the middle-density region ($0.2\lesssim \phi_{\rm cy}\lesssim 0.5$)
around the critical points (see Figs.~\ref{fig:snaps}(c) and (d)).
The bound membranes form an ellipsoidal shape, whereas the unbound membranes form a saddle shape between them.
In Figs.~\ref{fig:tubec1} and \ref{fig:tubec15}, the data of the phase-separated tubes are not plotted.
The homogeneous phases disappear for $3.2\lesssim f_{\rm ex}\sigma/k_{\rm B}T\lesssim 3.9$ at $C_0\sigma=0.075$ and $\mu_{\rm ms}/k_{\rm B}T=3$ 
(see Figs.~\ref{fig:snaps}(c) and Movie 2 provided in ESI
using the same parameter set with different initial conformations).
This beaded-necklace-like tube is also formed when the homogeneous tube becomes unstable
at the ends of the metastable state of the coexistence region (see Figs.~\ref{fig:snaps}(d)).
Membrane is ruptured when a neck region becomes too narrow.
The coexistence region is narrower in the simulation than in the theory.
The phase separation likely causes this reduction of the coexistence region.

Similar necklace-like membrane tubes with phase separation have been previously observed in the experiments on
three-component membranes~\cite{yana08,yana10}.
Moreover, membrane fission has been observed in the necked region~\cite{alla04}.
In their system, the difference in bending rigidity causes a neck-like shape
and the phase separation occurs even in a flat membrane.
In contrast, for the present system, the difference in the spontaneous curvature gives more dominant effects.

Unduloid-like deformations have been reported in tethered vesicles after fore release~\cite{tozz19}, 
as well as in tubular vesicles with polymer anchoring~\cite{tsaf01} and rolled membranes during detachment from a substrate~\cite{nogu19c}.
In contrast, phase separation can make cylindrical tubes unstable under strong forces.
This beaded-necklace-like shape may be expressed by a periodic combination of two constant-curvature surfaces discussed in Ref.~\citenum{gozd99} with an extension to include the deformation by the external force.
 
Although Fig.~\ref{fig:snaps} shows only necklace-like membranes of longer tubes,
the shorter tubes with $N=2400$ also show this structure with fewer periodicities (one or two circular bumps).
The simulation results exhibit no notable differences between these two tube sizes, 
except for slightly wider coexistence regions for the shorter tubes.

Here, we used a constant chemical potential and constant external force.
The same membrane shapes can be obtained with the ensemble of a constant number $N_{\rm p}$ of the proteins and/or constant tube length $L_{\rm cy}$,
when the condition is adjusted.
Note that for fixed $N_{\rm p}$ or $L_{\rm cy}$,
phase separation appears more often in the regions of the first-order transitions owing to the macroscopic phase separation.
For example, at $\phi_{\rm cy} \simeq 0.4$ and $f_{\rm ex}\sigma/k_{\rm B}T > 40$,
no solution exists in Figs.~\ref{fig:tubec1} and \ref{fig:tubec15}.
Thus, when $N_{\rm p}/N=0.4$, the membrane is separated into two regions of high and low protein densities.
This is similar to the gas--liquid coexistence in the $NVT$ ensemble~\cite{wata12}.

\section{Summary and discussions}\label{sec:sum}

We have studied the binding of curvature-inducing proteins onto the tethered vesicle.
Proteins exhibit an isotropic spontaneous curvature
such that they sense and generate the curvature of membranes.
For a completely unbound membrane,
the tube curvature and tube length are proportional to the force strength.
As the binding chemical potential increases,
the protein density sigmoidally increases in the spherical component of the vesicle.
In contrast, a discrete increase can also occur in the membrane tube,
accompanied by a change in the tube radius.
The force--density curve and force--tube-curvature curve are
reflection and point symmetric to the point, where the tube curvature 
equals to the sensing curvature, respectively. 
The approximation that neglects the tube volume for a small tube radius
well reproduces the results of the analysis for a finite volume.
Meshless simulations of membrane tubes were conducted to confirm
these theoretical results.
The results of the simulation and theory
for the homogeneous phases show very good agreement.
Additionally, in the simulation, beaded-necklace-like membrane tubes with phase separation are found
around the critical points.
Membrane deformation induces this microphase separation.

Based on these mean-field analyses,
we propose a method for estimating the bending rigidity change by protein binding.
The sensing curvature is obtained as the tube curvature at the maximum protein density $\phi_{\rm cy}^{\rm max}$ under variation in the tube radius.
The bending rigidity ratio is obtained from the slope of $d(1/R_{\rm cy})/df_{\rm ex}$ at this maximum protein density
as expressed in Eq.~(\ref{eq:fext3}).
 Our simulation suggest that not large density, $\phi_{\rm cy}^{\rm max} \lesssim 0.5$, should be used
for these estimations to avoid the influence of inter-protein interactions.
The maximum density can be varied by
the binding chemical potential $\mu$, which is a function of the buffer protein concentration $\rho$.
For a dilute solution, $\mu(\rho)=\mu(1) + k_{\rm B}T\ln(\rho)$.

In this study, we  consider only proteins that isotropically bend the membrane and have no preferred lateral direction.
The BAR superfamily proteins exhibit anisotropic spontaneous curvatures.
Previously, we have reported that such proteins show characteristic behavior in tubular membranes~\cite{nogu16a,nogu14,nogu15b,nogu19a}; 
the force--protein-curvature curve has a flat region at low curvature owing to the adjustment of the protein orientation,
and protein assembly induces  elliptic and polyhedral tube formations.
However, the mean-field theory of the isotropic spontaneous curvature has been used to analyze the experimental results on
 the binding of the BAR proteins to the tethered membranes~\cite{prev15}.
Recently, a mean-field theory for a nematic order coupled with protein bending energy was developed 
for a fixed-shaped membrane~\cite{tozz21} based on Nascimentos' liquid-crystal theory~\cite{nasc17}. 
The present method can be extended to  anisotropic proteins by including the orientational degree.
This is one of the directions for further studies.

\begin{acknowledgments}
This work was supported by JSPS KAKENHI Grant Number JP21K03481.
\end{acknowledgments}

\end{document}